\documentclass[12pt]{article}
\usepackage{geometry}                
\usepackage{graphicx}
\usepackage{amssymb}
\usepackage{cite}
\newcommand{\beq}{\begin{equation}}
\newcommand{\eeq}{\end{equation}}
\newcommand{\bea}{\begin{eqnarray}}
\newcommand{\eea}{\end{eqnarray}}

\title{
\textbf{Hadronic Light-by-Light Scattering in\\ the Muonium Hyperfine Splitting}\\[1cm]
\author{
\textbf{S.~G.~Karshenboim{\boldmath $^{a,b}$}, V.~A.~Shelyuto{\boldmath $^a$}, and A.~I.~Vainshtein{\boldmath $^{c,d}$}}\\
[4mm]
$^a$ \normalsize\emph{D. I. Mendeleev Institute for Metrology, St. Petersburg 190005, Russia}\\
$^b$ \normalsize\emph{Max-Planck-Institut f\"ur Quantenoptik, Garching 85748, Germany}\\
$^c$ \normalsize\emph{William I. Fine Theoretical Physics Institute, School of Physics and Astronomy,}\\
\normalsize\emph{University of Minnesota, Minneapolis, MN 55455, USA}\\
$^d$ \normalsize\emph{Theory Group, Physics Department, CERN, CH-1211 Geneva 23,
  Switzerland}
}}
\date{}
\begin{document}
\maketitle
 \thispagestyle{empty}

 \vspace{-12cm}

\begin{flushright}

FTPI-MINN-08/21\\
UMN-TH-2609/07\\
CERN-PH-TH/2008-126
\end{flushright}

\vspace{10cm}

\begin{abstract}
We consider an impact of hadronic light-by-light scattering on the
muonium hyperfine structure. A shift of the hyperfine interval
$\Delta \nu({\rm Mu}) _{\rm\tiny HLBL}$ is calculated with the
light-by-light scattering approximated by exchange of pseudoscalar
and pseudovector mesons. Constraints from the operator product
expansion in QCD are used to fix parameters of the model similar to
the one used earlier for the hadronic light-by-light scattering in
calculations of  the muon anomalous magnetic moment. The
pseudovector exchange is dominant in the resulting shift, $\Delta
\nu({\rm Mu})_{\rm\tiny HLBL}= -0.0065(10) \,\mbox{Hz}$.
Although the effect is tiny it is useful in understanding  the level of hadronic
uncertainties.
\end{abstract}

\newpage
\renewcommand{\thepage}{\arabic{page}}
\setcounter{page}{1}

\section{Introduction}

Pure leptonic objects, such as free electron and muon or leptonic
bound systems, positronium and muonium, are of specific interest
because they allow {\em ab initio\/} calculations with a high
accuracy. There is no effect of strong interactions in the leading
terms and in a number of terms in perturbative series. Still
hadronic effects enter through higher loops in electromagnetic
and electroweak interactions.

\begin{figure}[h]
\centerline{\includegraphics[width=3cm]{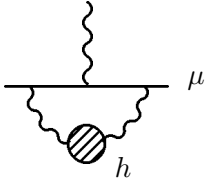}}
\vspace{-0.4cm}
\caption{\small The hadronic vacuum polarization contribution to $a_\mu$}
\label{amuVP}
\end{figure}

The most important leptonic property affected by hadronic effects is
the anomalous magnetic moment of a muon, $a_{\mu}=(g_{\mu}-2)/2$, where the main
hadronic contribution comes from the vacuum polarization (HVP), see
Fig.\,\ref{amuVP}.
The hadronic contribution is quite small $\Delta a_\mu({\rm
HVP})\simeq 7\times 10^{-8} \simeq 7\times 10^{-5}\,a_\mu$, but
nevertheless it is much larger than the experimental error in the
$a_{\mu}$ measurement \cite{lastexp},
\begin{equation}
a_{\mu}^{\rm exp}=116\;592\;080(63)\times 10^{-11}\,,
\end{equation}
 as well as the
uncertainty of the QED calculations \cite{kino} and  the
electroweak contribution. The HVP contribution is obtained
with sufficient accuracy by applying data from $e^+e^-$ annihilation into hadrons.

\begin{figure}[h]
\vskip -1cm
\centerline{\includegraphics[width=4cm]{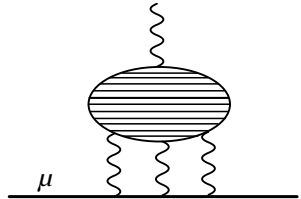}}
\caption{\small The light-by-light  scattering contribution to $a_\mu$}
\label{amulbl_1}
\end{figure}

At this level of accuracy one need to take into account higher order
hadronic effects and, in particular, the virtual light-by-light
scattering (HLBL), see Fig.\,\ref{amulbl_1}.
In contrast to HVP there is no direct experimental input for determining
 HLBL so one should rely on a theoretical model.

\begin{figure}[h]
\centerline{\includegraphics[width=9cm]{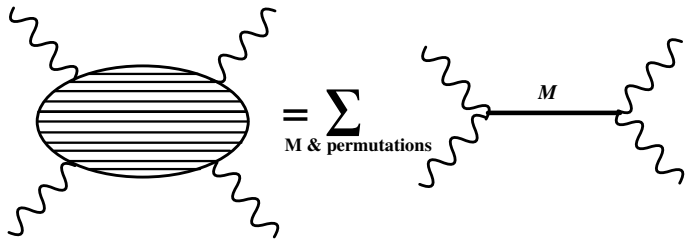}}
\caption{\small Meson exchanges in the light-by-light  scattering. Summation goes over
interchanges of photons and over $C$-even neutral mesons.}
\label{lbl_M}
\end{figure}

Two relevant theoretical parameters are the smallness of the chiral
symmetry breaking, $m_{\pi}^{2}/m_{\rho}^{2}\ll 1$, and the large
number of colors, $N_{c}\gg 1$. The first parameter enters a
powerlike, $1/m_{\pi}^{2}$ chiral enhancement for the charged pion
loop in HLBL while the large $N_{c}$ limit implies dominance of
meson exchanges, see Fig.\,\ref{lbl_M},
where mesons $M$ include
neutral pion and heavier $C$-even resonances.

In a number of papers dwelt on the problem it was shown that the
chirally enhanced two-pion contribution is significantly smaller
than the color enhanced one
\cite{Kinosh,Bijnens,Knecht,pion0,pion3}. The model for
light-by-light scattering developed in \cite{pion3} is based also on
QCD constraints which follow from operator product expansion at
large photon virtualities. Together with the neutral pion the
exchange of pseudovector mesons plays major role in the model.

\begin{figure}[h]
\centerline{\includegraphics[width=4cm]{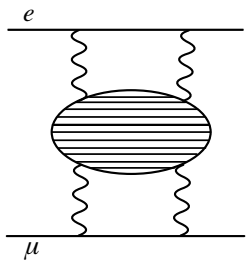}}
\vspace{-0.3cm}
\caption{\small The hadronic light-by-light scattering in the electron-muon
interaction}
\label{hfs}
\end{figure}

In the present paper we consider an impact of the hadronic light-by-light scattering
on another `pure leptonic' quantity, namely, to the muonium
hyperfine splitting (HFS), see Fig.\,\ref{hfs}.
 The hadronic effects in muonium are of
somewhat less practical importance since there has been no
experimental progress for years \cite{lampf}. However, the accuracy
in the former experiment was limited by statistics due to
low muon flux. At present, better muon sources are available, e.g.,
at the  Paul Scherrer Institut, and more accurate results
are in principle possible. To start preparation for a new
experiment one has to clearly understand the ultimate limit of the
theoretical accuracy.

\begin{figure}[h]
\centerline{\includegraphics[width=5cm]{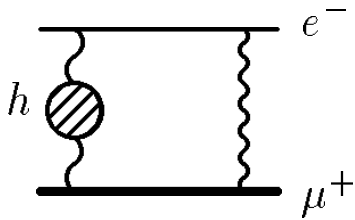}}
\caption{\small The characteristic diagram for the hadronic vacuum polarization
contribution to the muonium HFS interval}
\label{MuVP}
\end{figure}

In principle, pure QED calculations are {\em ab initio\/}
calculations and can be done with any accuracy (which does not mean
that they can be done easy---see, reviews \cite{report1,report2} for
the present status). However, the very involvement of the hadronic
effects sets a certain limit of accuracy. As well as in the case of
$a_\mu$ one has to calculate the HVP contribution in the leading
order \cite{hvp1Mu,hvp12Mu}, see Fig.\,\ref{MuVP},
and the next-to-leading term \cite{hvp2Mu,hvp12Mu}.

The HVP  contribution can be found from experimental data on the
$e^+e^-$ annihilation into hadrons. The HLBL contribution is of the
same order as the next-to-leading HVP contributions \cite{hvp12Mu}
and cannot be derived from existing scattering and annihilation
data. So we extend  the model of Ref.\cite{pion3} for the hadronic
light-by-light scattering to apply it to  the muonium HFS.

\begin{figure}[h]
\centerline{\includegraphics[width=3.5cm]{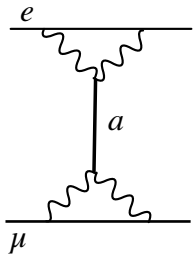}}
\vspace{-0.3cm}
\caption{\small The ``vertical'' exchange of pseudovector meson $a$ in $e\mu$ scattering}
\label{lbl_A}
\end{figure}

An interesting feature of this application is that the dominant contribution
comes from the ``vertical'' exchange by pseudovector mesons, see Fig.\,\ref{lbl_A}.
The reason for this dominance is that the pseudovector exchange
shown in  Fig.\,\ref{lbl_A} is the most relevant one  for the
spin-spin interaction of the electron and muon which  determines
 the HFS. By contrast, a similar exchange of a neutral pion vanishes
 in the scattering amplitude when the
electron and muon are at rest.

\begin{figure}[h]
\centerline{\includegraphics[width=6cm]{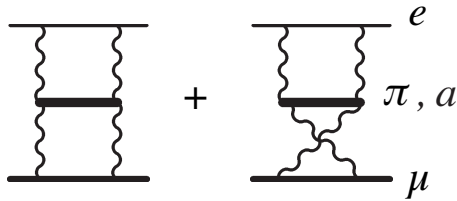}}
\vspace{-0.3cm}
\caption{\small The ``horizontal'' exchange of pseudoscalar $\pi$ and pseudovector
mesons $a$ in the $e\mu$ scattering}
\label{hfs-lbl_hor}
\end{figure}

The pion and pseudovector cross-channel ``horizontal'' exchange, see
Fig.\,\ref{hfs-lbl_hor},
are also accounted in the model. This contribution is numerically smaller
than the ``vertical'' one. Thus, the
situation in HFS is opposite to that for the $a_\mu$, where the pseudoscalar
exchanges dominate.

Another  interesting point is  that  the chirally enhanced charge
pion loop in the blob of  Fig.\,\ref{hfs} does not contribute to
spin-dependent part of the scattering amplitude. Indeed, the quantum
numbers $J^{P}$ of exchange should be $1^{+}$ as for pseudovector
mesons. However, such quantum numbers are not allowed  for the pair
$\pi^{+}\pi^{-}$.   Thus, in contrast to $a_{\mu}$, an ambiguous
charged pion loop does not enter the muonium HFS.

In the next section we introduce general expressions for the HLBL
effect in HFS.  In Sec. 3 we present calculations of the
pseudovector exchange, and in Sec. 4 we consider the pseudoscalar
exchange. In the last section we summarize the results.

\section{Generalities}

Let us start with some general expressions. The muonium HFS is
determined by the spin-dependent part of the
forward $e^{-}\mu^{+}$ scattering in the low-velocity limit. It is
convenient to start with the $e^{-}\mu^{-}$ amplitude and then make
the charge conjugation for the muon.

The spin-dependent part of the forward $e^{-}(p)+\mu^{-}(r)\to e^{-}(p)+\mu^{-}(r)$ scattering associated with HLBL can be presented as
\begin{equation}
M^{\rm spin}(e^{-}\,\mu^{-}\to e^{-}\,\mu^{-})=A \,\bar u^{(e)}\gamma^{\sigma}\gamma_{5} u^{(e)}\,
 \bar u^{(\mu)}\gamma_{\sigma}\gamma_{5} u^{(\mu)}\longrightarrow -4m_{e}m_{\mu}A\,\vec \sigma_{e}\,\vec \sigma_{\mu}\,,
 \label{ampl}
\end{equation}
where $u^{(e)}$, $u^{(\mu)}$ are Dirac spinors describing   the electron and muon  (we are using
relativistic normalization and units $\hbar=c=1$) and we took the nonrelativistic limit in the last
expression. The transition from $\mu^{-}$ to $\mu^{+}$ does not change the result because
of the positive $C$-parity of the axial current $ \bar u^{(\mu)}\gamma_{\sigma}\gamma_{5} u^{(\mu)}$.

The above  amplitude leads to the following addition in the $e^{-}\mu^{+}$ Hamiltonian,
\begin{equation}
\Delta H^{\rm HFS} = A \,\delta^{3}(\vec r)\,\vec \sigma_{e}\,\vec \sigma_{\mu} \,.
\end{equation}
This should be compared with the leading term for the $s$-wave HFS Hamiltonian,
\begin{equation}
H^{\rm HFS} = \frac{2\pi\alpha}{3m_{e}m_{\mu}} \,\delta^{3}(\vec r)\,\vec \sigma_{e}\,\vec \sigma_{\mu} \,,
\end{equation}
which gives for the HFS interval (Fermi energy $E_{F}$)
\begin{equation}
h \nu^{\rm HFS}=E_{F}=
\frac{8}{3}\frac{\alpha^4m_e^2}{m_\mu}\left(\frac{m_\mu}{m_e +m_\mu}\right)^3\simeq h\cdot 4.459\!\times\!10^{9}\;{\rm Hz}\,.
\end{equation}
The shift in the splitting due to HLBL is
\begin{equation}
h \Delta\nu_{\rm HLBL}=\Delta E_{\rm HLBL}=E_{F}\,\frac{3m_{e}m_{\mu}}{2\pi\alpha}\,A
\,.
\end{equation}

The amplitude $A$ is defined by the diagram in Fig.\,\ref{hfs},
\begin{equation}
A=\frac{4\alpha^{2}}{3}\int \frac{d^{4}k d^{4}q}{(2\pi)^{6}}\,\frac{M_{\mu\nu\mu'\nu'}(k,q)}
{(k^{2})^{3}q^{2}[(q^{2})^{2}-4(rq)^{2}]}
\,\epsilon^{\mu'\!\rho\mu\delta}k_{\rho}\epsilon^{\nu'\!\rho'\!\nu\delta'}q_{\rho'}
\Big (g_{\delta\delta'}-\frac{r_{\delta}r_{\delta'}}{m_{\mu}^{2}}\Big )
\,.
\label{master}
\end{equation}
Here $M_{\mu\nu\mu'\nu'}(k,q)$ is the the amplitude of the forward scattering of
two virtual photons,
\begin{equation}
\gamma^{*}(\mu,k)+\gamma^{*}(\nu,q)\to \gamma^{*}(\mu',k)+\gamma^{*}(\nu',q)\,,
\end{equation}
(we mark their polarization indices and momenta),  $r=\{m_{\mu},0\}$ is the 4-momentum
of the muon at rest, and we neglected by the electron mass. In many cases, in particular for
pseudovector exchanges, one can neglect by the muon mass as well. Then
the expression in Eq.\,({\ref{master}) simplifies further,
\begin{equation}
A=\alpha^{2}\!\int \frac{d^{4}k d^{4}q}{(2\pi)^{6}}\,\frac{M_{\mu\nu\mu'\nu'}(k,q)}
{(k^{2})^{3}(q^{2})^{3}}\,\epsilon^{\mu'\!\rho\mu\delta}k_{\rho}\,\epsilon^{\nu'\!\rho'\!\nu}_{\hskip 5mm \delta}
\,q_{\rho'}\,.
\label{master1}
\end{equation}

\section{Pseudovector exchange}
To calculate the ``vertical'' pseudovector exchange, see Fig.\,\ref{lbl_A}, let us start by introducing
the effective vertex for  lepton interaction with the  pseudovector meson $a$,
\begin{equation}
h_{a}\,a_{\rho} \,\bar l \gamma^{\rho} \gamma_{5}l\,, \qquad l=e,\,\mu\,,
\label{alep}
\end{equation}
where $h_{a}$ is the coupling constant and $a_{\rho}$ is the
polarization of the axial meson. Implying the same $h_{a}$ for the
electron and muon (corrections due to their mass difference are
small and can be accounted for ) we get the  pseudovector contribution
to the amplitude $A$ of
the forward $e\mu$ scattering, see Eq.\,(\ref{ampl}),
\begin{equation}
A^{\rm PV}_{\rm vert}=-\frac{h_{a}^{2}}{m_{a}^{2}}\,,
\label{apv}
\end{equation}
where $m_a$ is the  pseudovector  meson mass. Note the negative sign
which follows from unitarity (see Ref.\,\cite{Eides} for a detailed discussion of the sign).
Hence even before explicit calculation
we know that the  pseudovector exchange correction to HFS is negative.

\subsection{Coupling of pseudovector mesons to photons and leptons}
The next step is to calculate $h_a$. To fix the
$a\gamma^{*}\gamma^{*}$ vertex one can use that at large
virtualities the operator product expansion relates product of two
electromagnetic currents to the axial current \cite{Bj}, see also
\cite{pion3},
\begin{equation}
\label{jj}
\int \! {\rm d}^{4}x {\rm d}^{4}y \,{\rm e}^{-iq_{1}x-iq_{2}y}\,
T\left\{j_{\mu}(x) j_{\nu}(y)\right\}=\!
\int \! \!
{\rm d}^{4}z\,
{\rm e}^{iq_3 z}
\frac{2\epsilon_{\mu \nu \delta\rho}\,{\hat  q}^{\delta}}{{\hat q}^{2}}\,
j_{5}^{\rho}(z)+\cdots\,.
\end{equation}
Here
\begin{equation}
j^{\rho}_{5}=\bar q\, \hat Q^{2} \gamma^{\rho} \gamma_{5}\,q
\end{equation}
is
the axial current, where different flavors enter
 with weights proportional to squares of their
electric charges, $q_3 = q_1+q_2$ and
$\hat q=(q_{1}-q_{2})/2\approx q_{1}\approx -q_{2}\,$.

The $a\gamma^{*}\gamma^{*}$ vertex which satisfies this constraint at large $q^{2}$
and regular at small $q$ can be chosen in the form
\begin{equation}
\label{agg}
V_{\rho\mu\nu}a^{\rho}=\frac{ie^{2}\langle a| j^{\rho}_{5}(0)|0\rangle}{(q_{1}^{2}-m_{v}^{2})(q_{2}^{2}-m_{v}^{2})}\,\Big[q_{2}^{2}\epsilon_{\mu\nu\delta\rho}q_{1}^{\delta}+(q_{1}\leftrightarrow q_{2},\,\mu \leftrightarrow \nu)\Big] \,.
\end{equation}
Here $\langle a| j^{\rho}_{5}(0)|0\rangle$ is the matrix element between vacuum and the outgoing
axial meson with 4-momentum $q_{3}=q_{1}+q_{2}$ and polarization $a^{\rho}$ (photon momenta $q_{1}$ and $q_{2}$ momenta are taken as incoming). The form factor parameter $m_{v}$ is the mass of the  appropriate vector meson. Of course, this form of the
vertex is model-dependent. This refers not only to the above form of $q_{1}^{2}, q_{2}^{2}$ dependence but  also to choosing a particular structure, one of  three possible structures for the vertex. The choice (\ref{agg}) picks up the structure which survives in asymptotics.

The lepton interaction with $a$ can be calculated then from the triangle diagram (the upper and lower
blocks in Fig.\,\ref{lbl_A}). Taken all external momenta to be
vanishing and neglecting by lepton mass we get
\begin{eqnarray}
V&=&-2e^{4}\langle a| j^{\rho}_{5}(0)|0\rangle \epsilon_{\mu\nu\delta\rho}\!\int\! \frac{d^{4}q}{(2\pi)^{4}}\,\frac{q^{\delta}}{q^{2}(q^{2}-m_{v}^{2})^{2}} \,\bar l \gamma^{\mu}\,\frac{1}{\not \!  q}\,\gamma^{\nu}l\nonumber\\
&=&-\frac{3\alpha^{2}}{m_{v}^{2}}\,\langle a| j^{\rho}_{5}(0)|0\rangle\, \bar l \gamma_{\rho}\gamma_{5} l
 \,.
 \end{eqnarray}
 It gives the result for the pseudovector coupling to leptons  in terms of the vector mass $m_{v}$ and the matrix element of the axial current between the vacuum and  pseudovector meson,

Actually there are three electrically neutral pseudovector mesons $a^{(k)}$ which differ
by their features under
flavor SU(3). Therefore  it is convenient to present
 the axial current
$j_{5\rho}=\bar q\, \hat Q^{2} \gamma_{\rho} \gamma_{5}\,q$
as a linear combination of axial
currents with the same SU(3) quantum numbers as the mesons $a^{(k)}$. In particular, we can
introduce
the isovector, $ j^{(3)}_{5\rho} = \bar q\lambda_{3}\gamma_{\rho}\gamma_{5}q$, hypercharge,
$ j^{(8)}_{5\rho} = \bar q\lambda_{8}\gamma_{\rho}\gamma_{5}q$, and the SU(3)
singlet,
$ j^{(0)}_{5\rho} = \bar q \gamma_{\rho}\gamma_{5}q$, and write
\begin{equation}
j_{5\rho}=\sum_{k=3,8,0} \frac{{\rm Tr} [\lambda_{k}\hat Q^{2}]}{{\rm Tr} [\lambda_{k}^{2}]}\,j^{(k)}_{5\rho}\,,
 \end{equation}
where $\lambda_{0}$ is the unity matrix. Accounting for mixing of the hypercharge and singlet
pseudovector mesons is simply done by substituting $\lambda_{8}$ and $\lambda_{0}$
by appropriate linear combinations.

Thus, we get for the coupling $h_{a}^{(k)}$ of the meson $a^{(k)}$ to leptons (see Eq.\,(\ref{alep})
for definition)
\begin{equation}
h_{a}^{(k)}=-\frac{3\alpha^{2}}{m_{v}^{2}}\,\frac{{\rm Tr} [\lambda_{k}\hat Q^{2}]}{{\rm Tr} [\lambda_{k}^{2}]}\,f_{a}^{(k)}\,,
\label{hak}
\end{equation}
where $f_{a}^{(k)}$ is the coupling of the meson $a^{(k)}$ to the corresponding axial current,
\begin{equation}
\langle a^{(k)}|j^{(k)}_{5\rho}|0\rangle=f^{(k)}_{a}a_{\rho}\,.
\end{equation}

\subsection{Coupling of pseudovector mesons to axial currents}
The value of $f^{(k)}_{a}$ can be fixed from consideration of the
transition of the axial current, $j^{(k)}_{5\rho}$, into two
photons. We consider a special kinematics when one of those photons
is soft with momentum $k\to 0$ and polarization $\epsilon^{\nu}$ and
another is virtual, carrying the same momentum $q$ as the axial
current. The transition amplitude can be represented as
\begin{equation}
\label{trian}
T_{\rho\mu\nu}^{(k)}\epsilon^{\nu}
\!=i\,\langle 0| \!\int\! {\rm d}^4 z\, {\rm e}^{iqz}
T\{ j^{(k)}_{5\rho} (z)\,ej_{\mu} (0)\}|\gamma \rangle\,.
\end{equation}
Generically, as it is shown in  \cite{CMV},  the transition $T_{\rho\mu\nu}^{(k)}$ can be written
in terms of two
Lorentz invariant amplitudes, $w^{(k)}_{L}\! (q^{2})$ and  $w^{(k)}_{T}\!(q^{2})$,
\begin{equation}
\label{trian1}
T_{\rho\mu\nu}^{(k)}\epsilon^{\nu}
\!=\! -\frac{ie^{2} N_{c}{\rm Tr}[\lambda_{k}\hat Q^{2}]}{4\pi^{2}}\!
\left\{\!w^{(k)}_{L}\! (q^{2})\,q_{\rho}q^{\sigma}\!\tilde f_{\sigma\mu}\!+\!
w^{(k)}_{T}\! (q^{2})\!\left(\!-q^{2}\tilde f_{\mu\rho}\!+\!q_{\mu}q^{\sigma}\!\tilde f_{\sigma\rho}\!-\!
q_{\rho}q^{\sigma}\!\tilde f_{\sigma\mu}\right)\!\right\},
\label{LTw}
\end{equation}
where $\tilde f_{\mu\rho}=\epsilon_{\mu\rho\gamma\sigma}k^{\gamma}\epsilon^{\sigma}$.

In perturbation theory, $w^{(k)}_{L,T}(q^{2})$ are computed from triangle diagrams with
two vector currents and an axial current. For massless quarks, we have
\begin{equation}
w^{(k)}_{L}\! (q^{2})=2w^{(k)}_{T}\!(q^{2})=-\frac{2}{q^{2}}
 \,.
\end{equation}
An appearance of the longitudinal part for the axial current which classically is conserved
is a signal of the famous Adler-Bell-Jackiw axial anomaly \cite{ABJ}. The pole at $q^{2}=0$
in $w^{(k)}_{L}\! (q^{2})$ is associated with propagation of  massless Goldstone particles,
the pion in case of $w^{(3)}_{L}$.

There is no perturbative corrections to these functions in the
chiral limit. Moreover, the longitudinal functions  $w^{(3,8)}_{L}$
protected even against nonperturbative corrections. It is not the
case for transversal functions $w^{(k)}_{T}$ where the pole should
be shifted from zero to vector and  pseudovector masses. A
particular model which account for this shift suggested in
\cite{CMV}  has the form (in the chiral limit),
\begin{equation}
\label{wT}
w^{(k)}_{T}\!(q^{2})=\frac{1}{m_{a}^{2}-m_{v}^{2}}\left[\frac{m_{a}^{2}}{m_{v}^{2}-q^{2}}-
\frac{m_{v}^{2}}{m_{a}^{2}-q^{2}}\right],
\end{equation}
where $m_{a,v}$ denote masses of pseudovector and vector mesons
in the given channel $k$.

 Equation (\ref{trian}) implies the following expression for the residue of the pole at $q^{2}=m_{a}^{2}$, \begin{equation}
 \label{res}
\lim_{q^{2}\to m_{a}^{2}}(q^{2}- m_{a}^{2})\,T_{\rho\mu\nu}^{(k)}\,\epsilon^{\nu}
\!=
f^{(k)}_{a}V_{\rho\mu\nu}(q_{1}=q, q_{2}=k)\epsilon^{\nu}\,.
\end{equation}
Comparing this with the residue from Eqs.\.(\ref{trian1}) and (\ref{wT}) we get the result for $f^{(k)}_{a}$.
In particular for $f^{(3)}_{a}$ we have
\begin{equation}
\label{expf}
\Big[f^{(3)}_{a}\Big]^{2}=\frac{N_{c}m_{\rho}^{4}}{2\pi^{2}}
 \,.
 \end{equation}
An independent way to find $f^{(3)}_{a}$ is to use Weinberg's sum rules to relate it with the
$\rho$ coupling to electromagnetic current, $\langle \rho|j_{\mu}|0\rangle=(m_{\rho}^{2}/g_{\rho})\rho_{\mu}$,
\begin{equation}
\Big[f^{(3)}_{a}\Big]^{2}=\left(\frac{2m_{\rho}^{2}}{g_{\rho}}\right)^{2}
 \,.
 \label{fak}
 \end{equation}
Then Eq.\,(\ref{expf}) implies an interesting relation
\begin{equation}
\frac{g_{\rho}^{2}}{4\pi}=\frac{2\pi}{N_{c}}
\end{equation}
reasonably good phenomenologically. This can be also compared with the QCD sum rule
result \cite{SVZ}, $g_{\rho}^{2}/(4\pi)=2\pi/e$, where $e$, the base of natural logarithm, enters
instead of $N_{c}$ -- a pretty good approximation for $N_{c}=3$.

\subsection{Pseudovector exchange results}

Combining Eqs.\,(\ref{apv}), (\ref{hak}) and (\ref{fak})
we get for the exchange by the  isovector $a_{1}(1260)$ meson,
\begin{equation}
A^{a_{1}}_{\rm vert}=
-\frac{3}{8}\,\frac{\alpha^{4}}{\pi^{2}}\,\frac{1}{m_{a_{1}}^{2}}\,.
\end{equation}
For the isoscalar pseudovector mesons $f_{1}(1285)$ and $f_{1}^{*}(1420)$ we assume
the ÒidealÓ mixing, similar to $\omega$ and $\phi$. It means that  $f_{1}$ has the $(\bar u u+\bar d d)/\sqrt{2}$ structure and $f_{1}^{*}$ is $\bar s s$; this assumption is consistent with
experimental data for decays of these resonances. In terms of the relevant axial currents
the linear combinations $(\lambda_{0}/3)-(\lambda_{8}/\sqrt{3})$ and $(2\lambda_{0}/3)+(\lambda_{8}/\sqrt{3})$ enter correspondingly.
Then, similarly to the $a_{1}$ exchange, we get
\begin{equation}
A^{f_{1}}_{\rm vert}=
-\frac{25}{24}\,\frac{\alpha^{4}}{\pi^{2}}\,\frac{1}{m_{f_{1}}^{2}}\,,\qquad
A^{f_{1}^{*}}_{\rm vert}=
-\frac{1}{12}\,\frac{\alpha^{4}}{\pi^{2}}\,\frac{1}{m_{f_{1}^{*}}^{2}}
\,.
\end{equation}

Altogether the ``vertical'' exchange by pseudovector mesons
produces
\begin{eqnarray}
\Delta E^{\rm PV}_{\rm vert}&=&-\left(\frac{\alpha}{\pi}\right)^{3}\frac{m_{e}m_{\mu}}{m_{a_{1}}^{2}}\,E_{F}
\left[
\frac{9}{16}+\frac{25}{16}\,\frac{m_{a_{1}}^{2}}{m_{f_{1}}^{2}}
+\frac{1}{8}\,\frac{m_{a_{1}}^{2}}{m_{f_{1}^{*}}^{2}}
\right]\nonumber\\
&=&-\left(\frac{\alpha}{\pi}\right)^{3}\frac{m_{e}m_{\mu}}{m_{a_{1}}^{2}}\,E_{F}\cdot 2.16=h\cdot (-0.0041~{\rm Hz})
\,.
\end{eqnarray}

Now  let us add up the ``horizontal'' pseudovector exchanges shown
in Fig.\,\ref{hfs-lbl_hor}. In the limit of heavy pseudovector mass
this exchange differs from the ``vertical'' one just by averaging
over angles and constitutes 1/3 of the ``vertical'' exchange.
Accounting for the finite pseudovector mass we found an extra
to 1/3 suppression of the ``horizontal'' exchange by the factor
0.614,
\begin{equation}
\Delta E^{\rm PV}_{\rm
horiz}=-\left(\frac{\alpha}{\pi}\right)^{3}\frac{m_{e}m_{\mu}}{m_{a_{1}}^{2}}\,E_{F}
\cdot 2.16\,\frac{0.614}{3}=h\cdot (-0.000\,84~{\rm Hz}) \,.
\end{equation}
Thus, the total for pseudovector exchange is
\begin{equation}
\Delta E^{\rm PV}=-\left(\frac{\alpha}{\pi}\right)^{3}\frac{m_{e}m_{\mu}}{m_{a_{1}}^{2}}\,E_{F}\cdot 2.6
=h\cdot (-0.0049~{\rm Hz})\,.
\label{PVt}
\end{equation}
Note that we limit ourselves by exchanges of pseudovectors with the
lowest mass in each flavor channel. Exchanges by higher $1^{+}$
excitation contribute in the same direction but probably are
numerically suppressed.

\section{Pseudoscalar exchange}

Let us start with the $\pi^{0}\gamma^{*}\gamma^{*}$ vertex,
\begin{equation}
V_{\mu\nu}=c_{\pi\gamma\gamma}F_{\pi\gamma^{*}\gamma^{*}}(k^{2},q^{2})\,\epsilon_{\mu\nu\rho\sigma}k^{\rho}q^{\sigma}\,.
\end{equation}
Here $k$ and $q$ are photon momenta, $\mu$ and $\nu$ are their polarization indices,
the constant $c_{\pi\gamma\gamma}$ is fixed by the width of $\pi^{0}\to \gamma\gamma$ decay
and $F_{\pi\gamma^{*}\gamma^{*}}(k^{2},q^{2})$ is the form factor of the transition,
$F_{\pi\gamma^{*}\gamma^{*}}(0,0)=1$. Theoretical expression for $c_{\pi\gamma\gamma}$
\begin{equation}
c_{\pi\gamma\gamma}=\frac{\alpha N_{c}}{3\pi F_{\pi}}
\end{equation}
follows from the Adler-Bell-Jackiw anomaly \cite{ABJ}. Indeed, it could be read off from the residue
of the pole in the longitudinal part in Eq.\,(\ref{LTw}).

The pion exchange gives then the following expression for the forward scattering
of two virtual photons:
\begin{equation}
M_{\mu\nu\mu'\nu'}^{\rm pion}=c_{\pi\gamma\gamma}^{2}
\big[F_{\pi\gamma^{*}\gamma^{*}}(k^{2},q^{2})\big]^{2}
\,\frac{\epsilon_{\mu\nu\rho\sigma}k^{\rho}q^{\sigma}\epsilon_{\mu'\nu'\rho'\sigma'}k^{\rho'}q^{\sigma'}}
{m_{\pi}^{2}-(k+q)^{2}}+\Big(\mu\leftrightarrow \mu'\,,~k\to -k\Big)\,.
\end{equation}
The third permutation involving $k\leftrightarrow -q$ vanishes for
the forward scattering. It means an absence of the ``vertical''
exchange for the pion mentioned earlier once atomic momenta are
neglected.\footnote{The higher order contributions to muonium HFS
due to the ``vertical'' pion exchange together with additional photon
 are suppressed by extra small
factors such as $\alpha$ and $m_e/m_\pi$.\label{foo}}

Now we have to substitute $M_{\mu\nu\mu'\nu'}^{\rm pion}$ to
Eq.\,(\ref{master}) and integrate over $k$ and $q$. By power
counting at large momenta it is simple to see that in absence of the
form factor $F_{\pi\gamma^{*}\gamma^{*}}(k^{2},q^{2})$ the integral
logarithmically diverges. The form factor provides a convergence
above momenta of order of $m_{\rho}$, while its infrared convergence
is regulated by pion and muon masses. The $\ln(m_{\rho}/m_{\pi})$ term can be determined
analytically,
\begin{equation}
\label{log2-4}
\Delta E_{\pi}^{\rm log} = -\left(\frac{\alpha}{\pi}\right)^{3} \frac{m_{e}m_{\mu}}{(4\pi F_{\pi})^{2}}\, E_F \cdot
 \frac{9}{8} \,\ln{\frac{m_\rho}{m_{\pi}}}= h\cdot (- 0.0042\;\mbox{Hz})\,.
\end{equation}
For  numerical estimates we use $m_\pi= 135\;$MeV,
$m_\rho=775\;$MeV, $F_\pi=92\;$MeV. The logarithm is not that
big, $\ln(m_\rho/m_{\pi})=1.75$ so a numerical integration with a
certain model for the form factor is needed.

For the form factor
\begin{equation}
F_{\pi\gamma^{*}\gamma^{*}}(k^{2},q^{2})=
\frac{m_\rho^4}{(m_\rho^2-k^2)(m_\rho^2-q^2)}
\label{formf}
\end{equation}
numerical integration gives
\begin{equation}
\label{numpi}
\Delta E_{\pi}= -\left(\frac{\alpha}{\pi}\right)^{3} \frac{m_{e}m_{\mu}}{(4\pi F_{\pi})^{2}}\, E_F \cdot
 0.61= h\cdot (- 0.0014\;\mbox{Hz})\,.
\label{Dpion}
\end{equation}
The suppression of the logarithmic result can be approximated by
substitution
\begin{equation}
\ln{\frac{m_\rho}{m_{\pi}}}\to \left(\ln{\frac{m_\rho}{m_{\pi}}} -1.2\right)
\end{equation}
 in Eq.\,(\ref{log2-4}).

The result (\ref{Dpion}) can be compared with the earlier calculation  of the pion
contribution by  Faustov and Martynenko \cite{pion2}. They found $\Delta E_{\pi}= h\cdot (+ 0.0011\;\mbox{Hz})$ for the same form factor
(\ref{formf}). While the magnitude is close we differ in the sign.
Note that Faustov and Martynenko also considered change of HFS
due to effect of  HLBL pion exchange on $a_{\mu}$, the effect we are not considering.

Strictly speaking the form factor (\ref{formf}) violates the QCD
constraints. It follows from the OPE expansion (\ref{jj}) that at
$q^{2}=k^{2}$ the form factor should  decrease as $1/q^{2}$ at large
Euclidean momenta, not as $1/q^{4}$ as in Eq.\,(\ref{formf}). So we
made numerical integration with the form factor which satisfies the
above mentioned constraint as well as other theoretical and
experimental limitations \cite{Knecht} (see also
\cite{pion3}),
\begin{equation}
F_{\pi\gamma^{*}\gamma^{*}}(k^{2},q^{2})= \frac{ m_\rho^4M_2^4
-(4\pi^2F_{\pi}^2/N_c)\Bigl[q_1^2q_2^2(q_1^2+q_2^2)+h_2\cdot
q_1^2q_2^2+h_5\cdot(q_1^2+q_2^2)\Bigr]}{\bigl(q_1^2-m_\rho^2\bigr)
\bigl(q_1^2-M_2^2\bigr)
\bigl(q_2^2-m_\rho^2\bigr)\bigl(q_2^2-M_2^2\bigr)}\,,
\end{equation}
where $M_2=1465\; {\rm MeV}\,, ~h_5=6.93\; {\rm GeV}^4\,, ~h_2=-10
\;{\rm GeV}^2$.
 The result of integration turns out to be
very close to the one in Eq.\,(\ref{Dpion}), the difference is
insignificant.

Calculations for the other pseudoscalars, $\eta(547)$ and $\eta'(958)$, can be
done in a similar fashion. We use their experimental two-photon
width to determine the two-photon couplings and simple
vector-dominance form factors for off-shell photons
with $m_\rho=m_\omega=775~{\rm MeV}$ 
and $m_\phi=1020~{\rm MeV}$,
\begin{eqnarray}
F_{\eta\gamma^{*}\gamma^{*}}(k^{2},q^{2})&=&\frac{5}{3}\,
\frac{m_\rho^4}{(m_\rho^2-k^2)(m_\rho^2-q^2)}
-\frac{2}{3}\,
\frac{m_\phi^4}{(m_\phi^2-k^2)(m_\phi^2-q^2)}\,,
\nonumber \\
F_{\eta'\gamma^{*}\gamma^{*}}(k^{2},q^{2})&=&\frac{5}{6}\,
\frac{m_\rho^4}{(m_\rho^2-k^2)(m_\rho^2-q^2)}
+\frac{1}{6}\,
\frac{m_\phi^4}{(m_\phi^2-k^2)(m_\phi^2-q^2)}\,,
\end{eqnarray}
based on the octet and singlet quark structure of $\eta$ and $\eta'$.
The results of numerical integration are
\begin{eqnarray}
\label{numeta}
 \Delta E_{\rm \eta}&=&
-\left(\frac{\alpha}{\pi}\right)^{3} \frac{m_{e}m_{\mu}}{(4\pi
F_{\pi})^{2}}\, E_F \cdot
 0.063= h\cdot (- 0.000\,14\;\mbox{Hz})\,,
\nonumber \\[1mm]
 \Delta E_{\rm \eta^\prime}&=&
-\left(\frac{\alpha}{\pi}\right)^{3} \frac{m_{e}m_{\mu}}{(4\pi
F_{\pi})^{2}}\, E_F \cdot
 0.046= h\cdot (- 0.000\,10\;\mbox{Hz})\,.
\end{eqnarray}
This can be compared with  calculations by Faustov and Martynenko
\cite{pion2}, they obtained 0.0002 Hz for $\eta$ and 0.0001 Hz for
$\eta^\prime$. Again, we have a sign difference.

Thus, the total for pseudoscalar exchanges is
 \beq
 \label{Dps}
 \Delta E^{\rm PS}=
-\left(\frac{\alpha}{\pi}\right)^{3} \frac{m_{e}m_{\mu}}{(4\pi
F_{\pi})^{2}}\, E_F \cdot
 0.72= h\cdot (- 0.0016\;\mbox{Hz})\,.
\eeq

\section{Summary}

Collecting the results (\ref{PVt}), (\ref{Dps}) for pseudovector
and pseudoscalar exchanges we come
to
\begin{eqnarray}
\Delta E_{\rm HLBL}\!\!&=&\!-\left(\frac{\alpha}{\pi}\right)^{3}m_{e}m_{\mu}\,E_{F}\left[\frac{2.6} {m_{a_{1}}^{2}} +\frac{0.72}{(4\pi F_{\pi})^{2}}\right]\nonumber\\[3mm]
&=&\!h\cdot(-0.0049~{\rm Hz} -0.0016~{\rm Hz})=h\cdot(-0.0065~{\rm Hz}).
 \end{eqnarray}
The main contribution is due to pseudovector exchange, the
``vertical'' one in Fig.\,\ref{lbl_A}. It is fivefold larger
than ``horizontal'' pseudovector exchange and threefold larger than the ``horizontal'' pion
exchange.

What is the accuracy of the result? We mentioned in Introduction an absence of the charged
pion loop associated with chiral enhancement. This makes the result more reliable.
Looking on variations of parameters
such as coupling of the pseudovectors to axial currents we would estimate the uncertainty
of the model for the dominant pseudovector exchange as 10\%.
Staying on the conservative side we ascribe a total uncertainty to be about 25\%
of the pseudovector``vertical'' contribution, i.e., about 0.001 Hz.  Thus, our final result is
\[
h\Delta \nu_{\rm HLBL}=\Delta E_{\rm HLBL} =-0.0065(10) \;\mbox{Hz}\;.
\]

We see that the HLBL correction is tiny and rather unobservable.
However, it shows the level of limitations on theoretical accuracy
which comes from hadrons. In our study we also obtained a few relations
for couplings of pseudovector mesons to photons, leptons and axial currents
which can be applied to variety of processes.

\subsection*{Acknowledgments}

We gratefully acknowledge helpful discussions with Simon Eidelman.
VAS  is grateful to the Max-Planck-Institut f\"ur Quantenoptik for hospitality and support
of his visit during which this work was done. AIV thanks the CERN theory
group for support of his long-term stay. He also thanks the Galileo Galilei Institute for
Theoretical Physics for the hospitality and the INFN for partial support
during the completion of this work.

This work was supported in part by RFBR (grants \# 06-02-16156 and
\# 06-02-04018), by DFG (grant GZ 436 RUS 113/769/0-3) and  by DOE
grant DE-FG02-94ER408.

\end{document}